\documentclass[aps,pra,reprint,amssymb,longbibliography,preprintnumbers,twocolumn,amsmath,superscriptaddress,floatfix]{revtex4-2}
\usepackage{graphicx}
\usepackage{dcolumn}
\usepackage{bm}
\usepackage{physics}
\usepackage{orcidlink}

\usepackage{hyperref}
\usepackage{booktabs}
\usepackage[T1]{fontenc}
\usepackage[utf8]{inputenc}
\usepackage[normalem]{ulem}

\begin{document}

\title{Pulse-Duration Control of Subcycle Multiband Electron Dynamics Extends the High-Harmonic Cutoff in a Light-Driven Insulator}

\author{Hortense Allegre~\orcidlink{0009-0007-0325-6094}}
\thanks{These authors contributed equally.}
\email{\\Contact author: hortense.allegre20@imperial.ac.uk}
\affiliation{Blackett Laboratory, Physics Department, Imperial College London, Exhibition Road, London SW7 2AZ, UK}
\author{Simon V. B. Jensen~\orcidlink{0000-0002-6749-0978}}
\thanks{These authors contributed equally.}
\email{\\Contact author: simon.jensen@mpsd.mpg.de}
\affiliation{Max Planck Institute for the Structure and Dynamics of Matter and Center for Free-Electron Laser Science, Hamburg 22761, Germany}
\author{Joseph J. Broughton~\orcidlink{0000-0002-1468-0962}}
\affiliation{Blackett Laboratory, Physics Department, Imperial College London, Exhibition Road, London SW7 2AZ, UK}

\author{Tim Klee~\orcidlink{0009-0007-2556-9542}}
\affiliation{Blackett Laboratory, Physics Department, Imperial College London, Exhibition Road, London SW7 2AZ, UK}

\author{Yan Li~\orcidlink{0009-0003-5243-0694}}
\affiliation{Blackett Laboratory, Physics Department, Imperial College London, Exhibition Road, London SW7 2AZ, UK}

\author{Jon P. Marangos}
\affiliation{Blackett Laboratory, Physics Department, Imperial College London, Exhibition Road, London SW7 2AZ, UK}

\author{Nicolas Tancogne-Dejean~\orcidlink{0000-0003-1383-4824}}
\affiliation{Max Planck Institute for the Structure and Dynamics of Matter and Center for Free-Electron Laser Science, Hamburg 22761, Germany}
\affiliation{Research Center Future Energy Materials and Systems of the University Alliance Ruhr and Interdisciplinary Centre for Advanced Materials Simulation, Faculty of Physics and Astronomy, Ruhr University Bochum, Universitätsstraße 150, D-44801 Bochum, Germany}

\author{Angel Rubio~\orcidlink{0000-0003-2060-3151}}
\email{Contact author: angel.rubio@mpsd.mpg.de}
\affiliation{Max Planck Institute for the Structure and Dynamics of Matter and Center for Free-Electron Laser Science, Hamburg 22761, Germany}
\affiliation{Initiative for Computational Catalysis (ICC), The Flatiron Institute, New York, New York 10010, USA}

\author{John W. G. Tisch}
\affiliation{Blackett Laboratory, Physics Department, Imperial College London, Exhibition Road, London SW7 2AZ, UK}
\author{Mary R. Matthews}
\email{Contact author: m.matthews@imperial.ac.uk}
\affiliation{Blackett Laboratory, Physics Department, Imperial College London, Exhibition Road, London SW7 2AZ, UK}
\date{\today}

\begin{abstract}

We demonstrate pathway-selective control of extreme-ultraviolet high-harmonic generation by jointly tuning laser pulse duration ($5$ - $29$ fs) and intensity ($0.8$ – $74$ TW/cm$^2$). Many-cycle pulses at moderate intensities, $\sim6$~TW/cm$^2$, promote cumulative carrier transfer over successive optical cycles, progressively accessing higher conduction bands.
In contrast, few-cycle, high-intensity, $\sim22$~TW/cm$^2$, pulses drive subcycle multiband dynamics that reach $25$ - $50$ eV photon energies before decoherence can suppress coherent emission.
These results reveal pulse duration and intensity as decisive control knobs for high-harmonic emission, opening a route to band-structure-guided pulse design for higher energy extreme-ultraviolet light sources.

\end{abstract}

\maketitle

\def\thefootnote{*}\footnotetext{These authors contributed equally to this work.}

High-order harmonic generation (HHG) in solids has emerged as a powerful platform for exploring ultrafast electron dynamics in condensed matter systems~\cite{Ghimire2011,Goulielmakis2022, vanEssen2024TowardSolids}, for spectroscopic applications \cite{Lanin2017,Nourbakhsh2021,Uzan-Narovlansky2023,Wang2017, Uzan-Narovlansky2022ObservationSpectroscopy}, and potential innovations in compact coherent extreme-ultraviolet (XUV) sources. In contrast to gas-phase HHG, where the semiclassical three-step model provides an intuitive physical description~\cite{Corkum1993PlasmaIonization,PhysRevLett.70.1599}, HHG in solids arises from a complex interplay of intraband Bloch electron motion, interband polarization, and scattering processes~\cite{Vampa2017MergeScience}. Indeed, electrons in a solid are not accelerated as free but quasi-free particles, whose energy and momentum are intrinsically connected to the material band structure~\cite{Schubert2014Sub-cycleOscillations, Zuo2021NeighboringGeneration,Tancogne-Dejean2017b}. The emitted harmonic spectrum thus holds great potential for advancing our understanding of ultrafast light-driven electron dynamics by encoding energy and phase information about carrier acceleration within bands, interband transitions, and coherence between electronic states. This enables strong-field all-optical spectroscopic applications probing phenomena such as Bloch oscillations~\cite{Luu2015}, dynamical band dressing~\cite{Thorpe2023HighChannels, Koll2025,PhysRevLett.121.097402}, and field-driven population transfer~\cite{Lakhotia2020LaserSolids, Jia2017NonadiabaticSolids}. Consequently, solid-state HHG has been proposed and demonstrated for reconstructing band dispersions~\cite{parks2025full}, probing Berry curvature~\cite{Luu2018MeasurementSpectroscopy}, and tracking phase transitions~\cite{Bionta2021TrackingSpectroscopy}.

While the dependence of solid-state HHG on driving field strength~\cite{Wang2017}, wavelength~\cite{Liu2018, Wang2017}, and crystal orientation~\cite{You2017AnisotropicCrystals} has revealed characteristic crystalline symmetries and spectral plateau cut-off scaling relations, the role of the driving pulse duration remains comparatively less explored. Pulse duration directly controls the temporal window for electron acceleration, the buildup of interband coherence, the extent of intraband motion, and the cumulative influence of scattering and dephasing. 
Jointly varying pulse duration and intensity therefore provides a controlled means to disentangle transient subcycle dynamics from longer-timescale processes such as carrier relaxation and decoherence, offering direct insight into the microscopic mechanisms of HHG. 
Although carrier-envelope phase effects may further modulate subcycle dynamics, they are not expected to alter the general trends identified here. This control strategy is motivated by atomic HHG, where pulse duration has been shown to play a central role in selecting quantum-path contributions, optimizing phase matching, improving conversion efficiency, and enhancing harmonic yield~\cite{Westerberg2025}.

In solids, a more complex harmonic emission process is present: intraband motion, multiband effects, electron-phonon scattering, and excitation or correlation-induced dephasing, decoherence, and dissipation all influence electron dynamics, even on few-femtosecond timescales~\cite{PhysRevB.108.115433,Hohenleutner2015,Langer2016}. These effects may enhance or mitigate the influence of pulse duration on the resulting HHG spectra, and insights from gas-phase HHG might not directly translate.

Here, we present an experimental study of high-harmonic generation in MgO as a function of driving-pulse duration and intensity, with the aim of isolating the role of the driving-field temporal profile. Our results demonstrate that pulse duration control can selectively promote single-cycle and subcycle excitation pathways involving multiple electronic bands. These findings establish pulse duration as a central control parameter in solid-state HHG, enabling manipulation of temporal coherence, spectral extension, and ultrafast multiband electron dynamics, with direct implications for coherent XUV light-source technologies.

\begin{figure}
\includegraphics[width=\columnwidth]{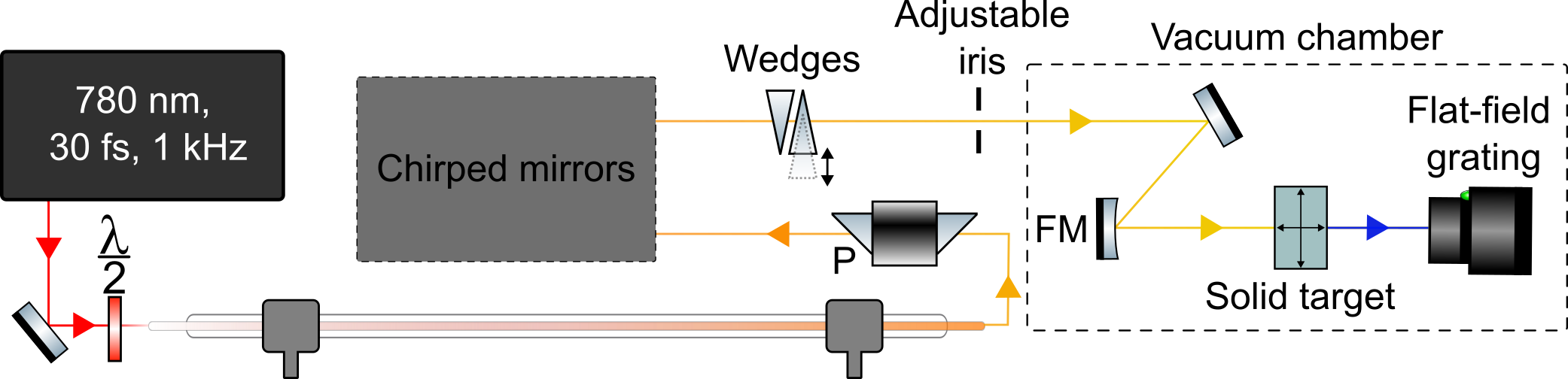}
\caption{\label{fig:setup} Schematic of the laser system: the laser pulse is spectrally broadened using an Ar-filled hollow-core fibre with pressure varied between $0$ and~$1$~bar. The pulse is temporally compressed by chirped mirrors, with the subsequent wedges allowing for optimisation. FM is a focusing mirror, $\lambda/2$ is a half-waveplate, and P is a linear polariser.}
\end{figure}

We use a Ti:Sapphire laser centred at $782$ nm ($1.59$ eV) to drive harmonics in MgO ((001), $100$ \textmu m thick). The laser is linearly polarised, applied at normal incidence in transmission geometry with polarization aligned with the $\Gamma-X$ direction, and with intensity control using a combination of a half-waveplate and a linear polariser~\cite{Broughton2025} as illustrated in Fig.~\ref{fig:setup}.
The pulse duration is tuned using a statically filled hollow-core fibre pulse compression system where varying Ar pressure controls the self-phase modulation~\cite{Okell2013}. A set of ten double-angled chirped mirrors, imparting a total group delay dispersion of $-360$ fs$^\text{2}$, imposes a near transform-limited pulse compression, and wedges allow fine tuning of the pulse dispersion. Lastly, an adjustable iris controls the beam size to improve signal-to-noise ratio without material damage.

The pulse is focused on the MgO sample in a vacuum chamber with a $50$ cm focal length mirror, and the generated harmonics are collected by a flat-field spectrometer as in Ref.~\cite{Allegre2025}. Pulse duration is measured with a d-scan (\textit{Sphere}) for the $5$ fs pulse and an SHG-FROG for the longer pulses. 
Parameters for all driving pulses are summarised in Table~\ref{table:pulses_parameters}, with pulse reconstructions in the supplementary information (S.I.) Sec.~I~\cite{SI}. Pulse duration measurements after the MgO sample displayed no significant change in the temporal and spatial driving laser profiles, verifying that the signatures and conclusions of this work are unlikely to arise from dispersion or nonlinear propagation effects. We note that the optical setup imposes different constraints on the intensity regime spanned by each pulse. Importantly, shorter pulses allow access to higher intensities whilst remaining below the material damage threshold, a property which will become key to opening different emission channels.

\begin{table}[h]
\begin{center}
\resizebox{0.5\textwidth}{!}{\begin{tabular}{l | c c c}
\toprule[1 pt]
Pulse duration $\tau$ (fs) & 29 & 17 & 5 \\ [0.5ex]
\bottomrule[1 pt]
Hollow-core fibre Ar pressure (bar) & 0.0 & 0.4 & 1.0 \\
Center wavelength (nm) & 782 & 766 & 696 \\
Pulse energy (\textmu J) & 4$-$56 & 3$-$45 &  2$-$58 \\
Beam radius ($1/e^2)$ \textit{w} (\textmu m) & 108 & 143 & 100 \\
Intensity (TW/cm$^\text{2}$) & 0.8$-$10 & 0.5$-$7.7 & 2.7$-$74\\
\bottomrule[1pt]
\end{tabular}}
\caption{Laser parameters for all applied driving laser fields.}
\label{table:pulses_parameters}
\end{center}
\end{table}
\nocite{PhysRevA.81.021403,Hummert_2020,PhysRevA.109.063109,jackson_classical_1999,PhysRevA.81.021403,10.1063/1.5142502,Tancogne-Dejean2017,PhysRevB.58.3641,PhysRevB.13.5188,PhysRevLett.102.226401,PhysRevB.87.075121,WHITED19731903,PhysRevA.98.023415,Silva2018,doi:10.1021/acsphotonics.9b00019,Tancogne-Dejean2017,Nourbakhsh2021,Uzan2020}

\begin{figure}
\includegraphics[width=\columnwidth]{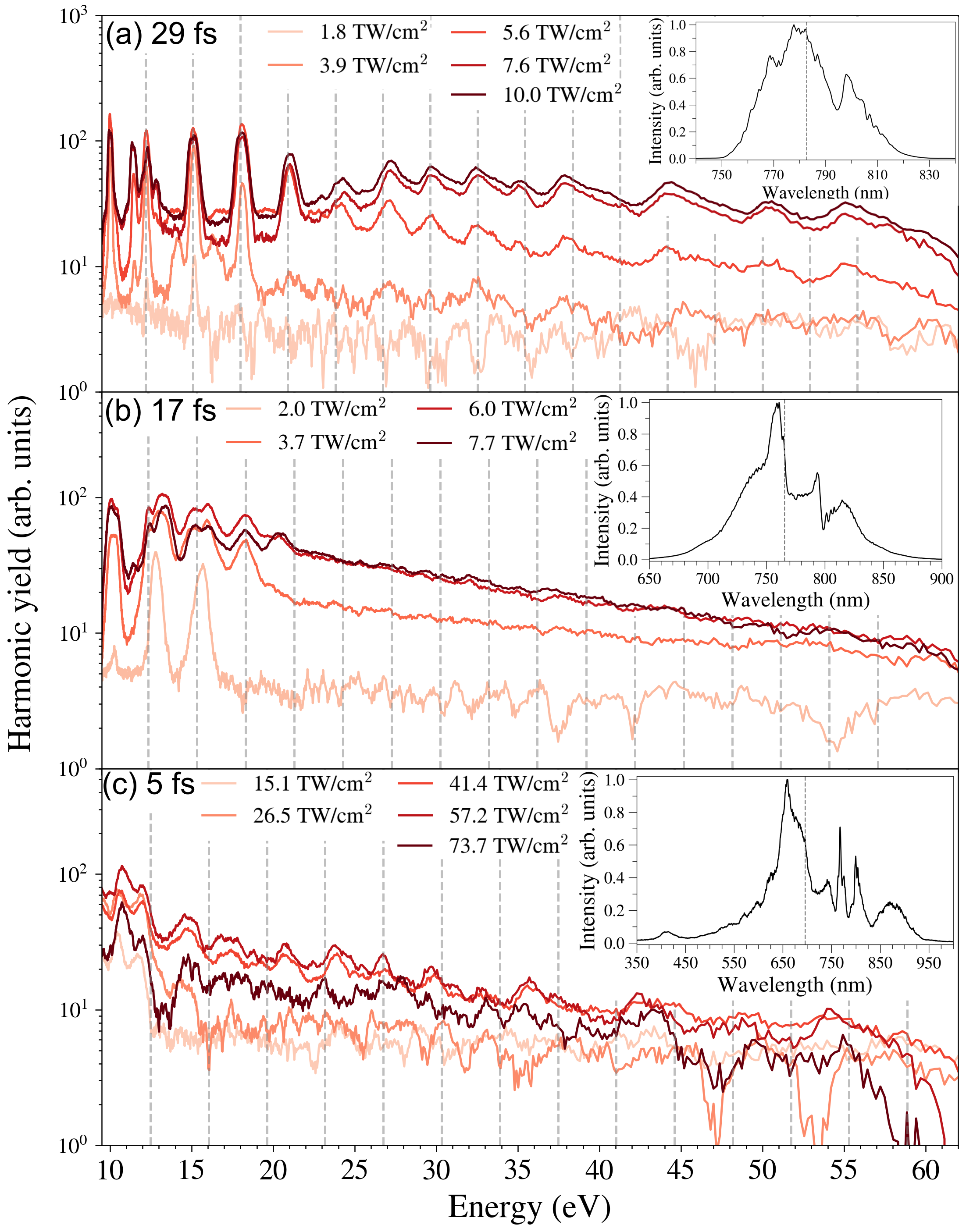}
\caption{\label{fig:spectrum_scan_duration}
Experimental high-order harmonic spectra from MgO driven by the pulses described in Table~\ref{table:pulses_parameters}~(a) $29$ fs, (b) $17$ fs, and (c) $5$ fs. The dashed lines in the harmonic spectra map the multiples of the central frequency stated in Table~\ref{table:pulses_parameters}. Inset: spectrum of the fundamental driving pulse, with dashed line corresponding to the central wavelength of each spectrum. }
\end{figure}

We present in Fig.~\ref{fig:spectrum_scan_duration} the measured harmonic spectra above the 7$^\text{th}$ harmonic, where a varying harmonic cutoff is obtained as a function of pulse intensity and duration.
Considering first the low intensity regime of $1.8$ - $5.6$~TW/cm$^2$ in Fig.~\ref{fig:spectrum_scan_duration}~(a), we observe an increase of both the low-order harmonic yields and the harmonic cutoff energy with driving field intensity as expected. However, once the driving intensity exceeds $\sim 6$ TW/cm$^\text{2}$, the harmonic cutoff shifts rapidly, introducing an additional plateau of high-energy harmonics from $25$ eV and towards $50$ eV~\cite{Allegre2025, wu2016multilevel, You2017a, Ndabashimiye2016}, consisting of relatively spectrally broadened harmonic peaks. 
As will be outlined later, the threshold intensity for generating the high-energy plateau matches a channel opening for electron transfer towards higher conduction bands during several laser cycles. Such a multi-step accumulative process impedes the yield of the low-energy plateau harmonics, and its reduced temporal window of emission explains the relatively broadened peak shapes in the high-energy plateau.
Counter-intuitively, the high-energy plateau disappears with decreasing $17$ fs pulse duration in Fig.~\ref{fig:spectrum_scan_duration}~(b) to later re-emerge for even shorter $5$ fs pulses in Fig.~\ref{fig:spectrum_scan_duration}~(c), but upon re-emerging, the first and second plateaus exhibit similar peak widths. We will attribute such reappearance to the opening of a different emission channel, whose subcycle nature grants relatively narrow peak widths similar to the lower plateau. 

\begin{figure}
\includegraphics[width=\columnwidth]{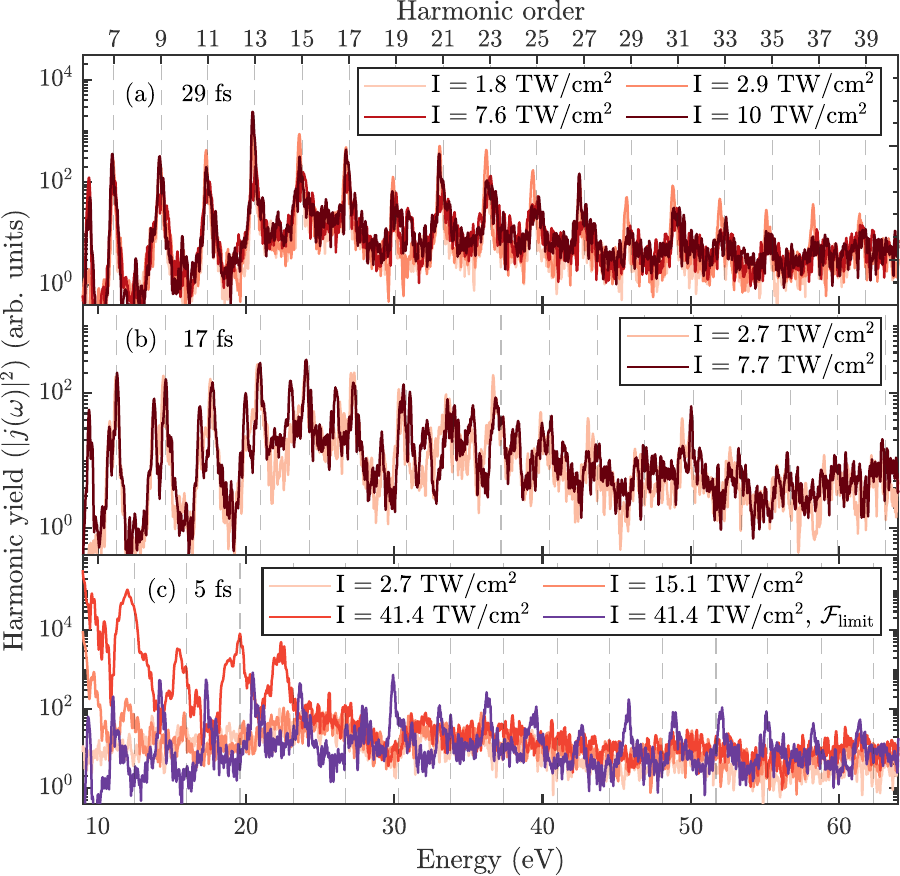}
\caption{\label{fig:theory}
Simulated high-order harmonic spectra from MgO driven by extracted experimental electric field traces of (a) $29$ fs, (b) $17$ fs, and (c) $5$ fs duration, using characteristic intensities of Fig.~\ref{fig:spectrum_scan_duration}. Panel (c) additionally compares towards a Fourier-limited ($\mathcal{F}_{\text{limit}}$) $780$ nm driving field. Dashed lines indicate harmonics of the extracted driving frequency.
}
\end{figure}

To clarify the role of the pulse duration, we consider an intra- and intercycle analysis, generalizing ideas of Refs.~\cite{PhysRevA.109.063109,PhysRevA.81.021403}. We assume the HHG process replicates periodicity during each of $N_c$ optical cycles of the driving field with frequency $\omega_L$.
Under such approximations, see S.I.~Sec.~II~\cite{SI}, the harmonic spectrum factorises into an intra- and intercycle component ${\rm Yield}(\Omega) \propto \abs{I^{{\rm intra}}(\Omega)}^2\abs{I^{{\rm inter}}(\Omega)}^2$. The intracycle component contains material-specific characteristics such as the traditional harmonic cutoff, while the intercycle term provides signatures related to the duration of the light-matter interaction, confining harmonic peaks to integer multiples of $\omega_L$, with yield $\sim N_c^2$ and full-width-at-half-maximum (FWHM) $\propto (N_c^2-1)^{-1/2}$. Keeping the intensity constant and compressing the driving pulse, i.e., reducing the number of laser cycles $N_c$, causes the harmonic peaks to decrease and widen, as observed in Fig.~\ref{fig:spectrum_scan_duration}~(a-c) and illustrated in S.I.~Fig.~2~\cite{SI}. In the experiment, additionally broadening can be attributed to the near transform-limited pulse duration, see the insets in Fig.~\ref{fig:spectrum_scan_duration}.

Since the spectral confinement of harmonic peaks is diminished by pulse compression, the $25$ - $50$ eV high-photon-energy plateau vanishes in Fig.~\ref{fig:spectrum_scan_duration}~(b) despite the driving intensities exceeding the $\sim 6$ TW/cm$^\text{2}$ threshold, where the high-energy plateau arose in (a). 
In comparison, the cutoff photon energy extends from the $11^\text{th}$ order at $\sim 17$ eV in Fig.~\ref{fig:spectrum_scan_duration}~(b) towards the $35^\text{th}$ order at $\sim 55$ eV of (a), an observation that highlights the sensitivity of the cutoff energy towards laser pulse duration.
In Fig.~\ref{fig:spectrum_scan_duration}~(c), we observe that the above $20$ eV harmonic peaks are recovered, however, only upon exceeding a higher intensity threshold of $\sim 30$ TW/cm$^\text{2}$. 
This finding highlights the complex role of performing compression of the driving pulse duration for HHG: while the associated increase in peak intensity can extend the cutoff, the accompanying increase in bandwidth can reduce the spectral clarity of the harmonic peaks. Accessing higher-energy emission, therefore, requires controlled pulse compression that balances the gain in peak intensity against spectral broadening, while remaining below the damage threshold.
Comparing Figs.~\ref{fig:spectrum_scan_duration} (a) and (c), the high-energy plateau appears less broadened than the lower-energy plateau. This suggests that the high-energy electron transitions responsible for the emission in Fig.~\ref{fig:spectrum_scan_duration}~(c) occur on faster timescales than those in Fig.~\ref{fig:spectrum_scan_duration}~(a), allowing them to accumulate a comparable degree of intercycle interference and thus produce a similarly well-resolved spectral response, for further details, see S.I.~Sec.~II~\cite{SI}.

To investigate the microscopic origin of the effects reported in Fig.~\ref{fig:spectrum_scan_duration}, we employ first-principles time-dependent density functional theory using the Octopus code~\cite{10.1063/1.5142502} (details in S.I.~Sec.~III~\cite{SI}). 
The microscopic electron response is coupled to macroscopic Maxwell's equations to obtain the correct field intensity in matter.
Figure~\ref{fig:theory} presents \textit{ab initio} simulations using the experimentally retrieved pulse shapes, as described in Sec.~I of the SI~\cite{SI}. 
The simulations reproduce the observed trend of harmonic broadening with decreasing pulse duration: in Fig.~\ref{fig:theory}~(a), the high-energy harmonics remain well defined across the full plateau, whereas in Fig.~\ref{fig:theory}~(c) this spectral clarity is retained mainly at photon energies below $30$ eV. Fig.~\ref{fig:theory}~(b), by contrast, exhibits pronounced peak structuring throughout, as multiple frequencies interplay and substantially reduce the overall spectral resolution.
Comparison with an idealized $5$ fs Fourier-limited pulse, shown by the blue curve in Fig.~\ref{fig:theory}~(c), indicates that the harmonic cutoff can be extended by further optimizing the bandwidth during pulse compression to retain spectral clarity across the full plateau.
The only feature not fully captured in simulations is the relative intensity scaling of the two plateaus in Fig.~\ref{fig:theory}~(a) compared with Fig.~\ref{fig:spectrum_scan_duration}~(a), likely due to the demanding convergence requirements and the absence of decoherence from electron–electron scattering in the calculations. 
Carrier-envelope-phase averaging may also contribute to the remaining differences between theory and experiment. Overall, the \textit{ab initio} simulations confirm that the experimental trends arise from microscopic electron dynamics governed by the interplay between pulse intensity and duration, rather than macroscopic propagation effects.

\begin{figure}
\includegraphics[width=\columnwidth]{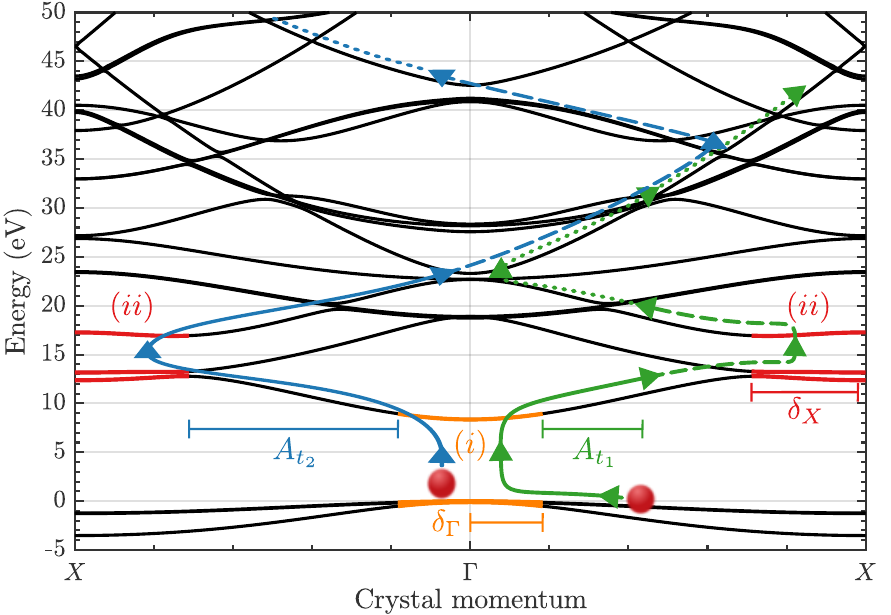}
\caption{\label{fig:theory2}
Calculated MgO band structure along the $\Gamma-X$ symmetry axis. Regions $(i)$ and $(ii)$ of radius $\delta_j$\, around symmetry points $j = \lbrace\Gamma,X\rbrace$, where interband transitions are most probable, appear in orange and red. Characteristic electron trajectories are given with threshold vector potentials $A_{t_1}$ (green arrow) and $A_{t_2}$ (blue arrow). The full, dashed, and dotted lines denote areas reachable during a half, a full, and two optical cycles, for the green trajectory reaching energies of, respectively, $\sim 10$ eV, $\sim 20$ eV, and $\sim 50$ eV.
}
\end{figure}

To understand the mechanisms responsible for the high-energy plateau in Figs.~\ref{fig:spectrum_scan_duration}~(a) and (c), we extend ideas of Ref.~\cite{PhysRevA.102.033105}, which identifies momentum space trajectories associated with high-energy plateaus for a chain model system.
For the MgO crystal, a limiting factor for generating high-energy harmonics is whether electrons can access band structure points where multiphoton absorption or Zener tunnelling towards higher bands occurs most efficiently. The initial excitation step (i), transferring carriers from the valence to the first conduction band, occurs predominantly around the $\Gamma$ point, see Fig.~\ref{fig:theory2}. Hereafter, efficient carrier transfer towards the second and third conduction bands occurs optimally near the $X$ point at step (ii). From the third conduction band, electrons are now able to access recombination energies beyond $20$ eV. 
Due to the laser bandwidth, the electronic band dispersion and its transition elements, excitations most efficiently occur across some $\delta_k$ around the high-symmetry points. Following Ref.~\cite{PhysRevA.102.033105}, we estimate the range of such $\delta_k \approx \sqrt{m^* \omega_L}$ by comparing the energy scale of the laser frequency $\omega_L$ with the dispersion defined by the reduced effective masses $m^*$ of the critical points. The extracted masses $m^* (\Gamma) = 0.36$, $m^* (X) = 0.90$ and associated ranges $\delta_\Gamma = 0.05\pi$ a.u.$^{-1}$ and $\delta_X = 0.07\pi$ a.u.$^{-1}$ are sketched in Fig.~\ref{fig:theory2} and corroborate the extracted transition dipole matrix elements, see S.I.~Sec.~III~\cite{SI}.

\begin{figure}
\includegraphics[width=\columnwidth]{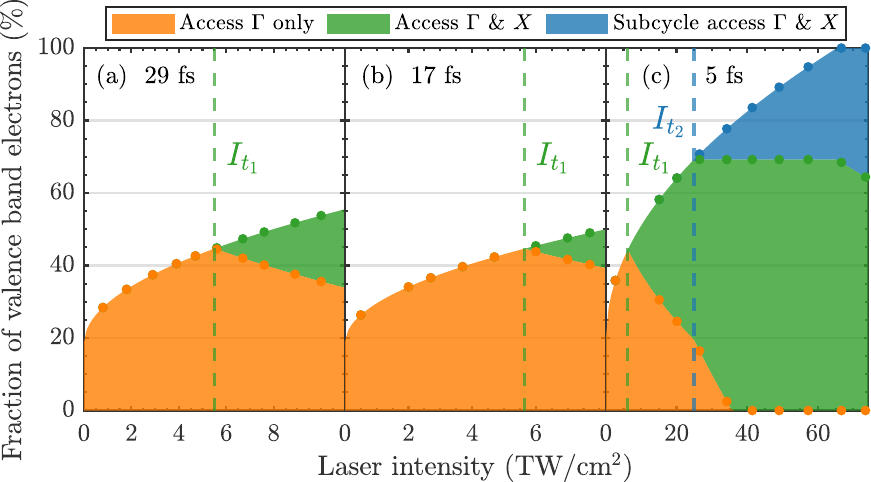}
\caption{\label{fig:theory3} Fraction of initial valence band crystal momenta $k_0$, that through the acceleration theorem can access $\Gamma \pm \delta_\Gamma$ only, or $\Gamma \pm \delta_\Gamma$ \& $X\pm \delta_X$ within a single or a half-laser cycle. Vertical lines denote the respective threshold intensities $I_{t_1}$ \& $I_{t_2}$ with associated characteristic trajectories sketched in~Fig.~\ref{fig:theory2}.}
\end{figure}

At low intensities, the harmonic cutoff energy increases with electric field strength $E$ or vector potential $A$, see Fig.~\ref{fig:spectrum_scan_duration}~(a) below $6$ TW/cm$^\text{2}$. Here, an electron excited at $(i)$ climbs the first conduction band via the acceleration theorem $k(t) = k_0 + A(t)$, recombining with energies determined by the energy gap at $k(t)$~\cite{Ghimire2011,Luu2015,PhysRevB.91.064302}. For higher intensities with vector potential exceeding the threshold $A_{t_1}$, an electron originating with a finite $k_0$ might pass by the region $(i)$ near $\Gamma$ within half of an optical cycle and by region $(ii)$ near $X$ within a subsequent half-cycle, as sketched with the green trajectory in Fig.~\ref{fig:theory2}. 
For the face-centered cubic lattice, this threshold vector potential is $A_{t_1} = \left( 2\pi/a-\delta_\Gamma - \delta_X \right)/2$ with lattice constant $a=4.212$ \AA, corresponding to a threshold intensity $I_{t_1} \approx 5.5$ TW/cm$^\text{2}$. 
This threshold matches the onset of the high-energy plateau in Fig.~\ref{fig:spectrum_scan_duration}~(a), and is also consistent with time-frequency analysis in S.I.~Sec.~IV~\cite{SI}. Although $A_{t_1}$ inherits a $\sqrt{\omega_L}$-dependence via $\delta_\Gamma$ and $\delta_X$, the dominant $2\pi/a$ term limits the resulting variation in $I_{t_1}$ to $\sim 12 \%$, over the experimental wavelength range. 

Since the high-energy plateau in Fig.~\ref{fig:spectrum_scan_duration}~(a) requires an accumulative multicycle carrier transfer to higher conduction bands compared to the single-cycle processes of the first plateau, it undergoes a greater reduction and broadening upon pulse compression, as observed in Fig.~\ref{fig:spectrum_scan_duration}~(b). Recovering this high-energy plateau requires electrons accessing both $\Gamma$ and $X$ to allow transitions above $20$~eV on a subcycle timescale. Such electron dynamics, sketched with blue in Fig.~\ref{fig:theory2}, require a vector potential of $A_{t_2} = 2\pi/a - \delta_\Gamma - \delta_X = 2 A_{t_1}$ corresponding to a threshold intensity $I_{t_2} = 4I_{t_1} \approx 22$ TW/cm$^\text{2}$. 
Such intensities are inaccessible with long laser pulses without exceeding the damage threshold, which is why this mechanism has received little attention compared with interpretations based on band climbing or cascaded excitation. Ultrashort pulses, however, can reach these intensities before damage occurs, enabling subcycle multiband excitations without relying on the slower, more decoherence-prone process of multicycle carrier accumulation. As shown in Fig.~\ref{fig:spectrum_scan_duration}~(c), once this intensity threshold is exceeded, the high-energy plateau reappears with relatively narrow harmonic peaks. These features originate from subcycle electron transitions, consistent with the time-frequency analysis in Sec.~IV of the SI~\cite{SI}.

To elucidate the microscopic mechanism, Fig~\ref{fig:theory3} shows the ratio of the initial electron wavevectors $k_0$ that can access the critical band structure points $(i)$ and $(ii)$ following the acceleration theorem. For the $29$ fs pulse, exceeding the threshold intensity $I_{t_1} \approx 5.5$ TW/cm$^\text{2}$ allows a fraction of the electrons to reach the $\Gamma$ and $X$ points within a single optical cycle, where interband tunnelling enables multiband trajectories akin to the green trajectory in Fig.~\ref{fig:theory2}.
If assuming that interband transitions are likely, then reaching $I_{t_1}$ will deplete the electron population remaining within the first conduction band. This depletion is reflected by the scaling relations of Fig.~\ref{fig:spectrum_scan_duration}~(a), where the high-energy plateau appears alongside a reduction of the lower-energy harmonics. For the shortened $17$ fs pulse, the intensity threshold of $I_{t_1}$ is also slightly exceeded in the experiment, but the high-energy plateau remains unseen in Fig.~\ref{fig:spectrum_scan_duration}~(b) due to spectral broadening. For the shortest $5$ fs pulse, it becomes possible for electrons to follow the blue trajectory of Fig.~\ref{fig:theory2} around $I_{t_2} \approx 22$ TW/cm$^\text{2}$, enabling subcycle electron transfer across the band structure and granting a reappearance of the high-energy plateau in Fig.~\ref{fig:spectrum_scan_duration}~(c). Comparative measurements and analysis along the $\Gamma - K$ direction are given in S.I.~Sec.~V~\cite{SI}, where larger momentum space distances and reduced effective masses increase intensity thresholds and impede the harmonic yield.

In conclusion, we have shown that by tuning the laser pulse duration, we can coherently control the ultrafast electron response and open subcycle multiband electron excitation channels, unreachable for longer pulses due to the damage threshold limit. When shortening the timescale associated with high-energy electron recombinations, we can further reduce the impact of the dephasing mechanisms that usually mitigate the electronic coherence and recombination efficiency. We therefore foresee that the driving-laser pulse duration will provide a powerful control parameter for enhancing harmonic emission in solid HHG experiments. By outlining the governing principles behind pioneering harmonic generation into the $50$ eV regime, we also open the door to enable next-generation compact solid-state XUV sources for attosecond spectroscopy and photonic technologies.

\begin{acknowledgments}
The technical support of S. Parker and A. Gregory is gratefully acknowledged. The experimental research was supported by the Royal Society URF \textbackslash R1\textbackslash191759, the Engineering and Physical Sciences Research Council (EPSRC), AWE and the Defence Science Technology Laboratory (DSTL). S.V.B.J. acknowledges support from the Alexander von Humboldt Foundation. This work was supported by the European Research Council (ERC-2024-SyG-101167294; UnMySt), the Cluster of Excellence Advanced Imaging of Matter (AIM), Grupos Consolidados y Alto Rendimiento UPV/EHU, Gobierno Vasco (IT1453-22). We acknowledge support from the Max Planck-New York City Center for Non-Equilibrium Quantum Phenomena. The Flatiron Institute is a division of the Simons Foundation.
\end{acknowledgments}

\bibliography{references}

\clearpage
\onecolumngrid
\begin{center}
\includegraphics[page=1,trim=1.9cm 1.8cm 1.8cm 1.8cm, clip, width=1\textwidth]{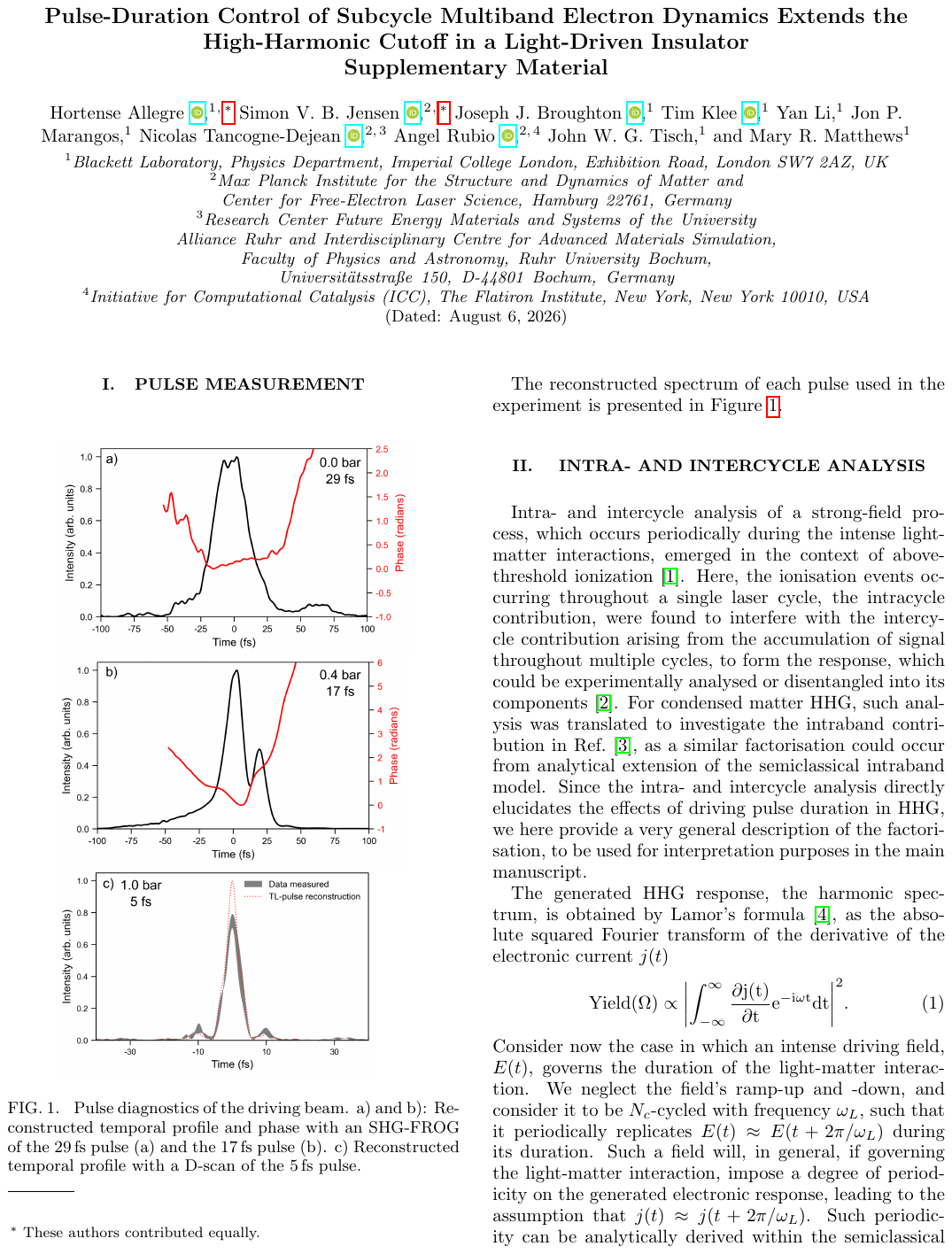}
\clearpage
\includegraphics[page=2,trim=1.9cm 1.8cm 1.8cm 1.8cm, clip, width=1\textwidth]{SI.pdf}
\clearpage
\includegraphics[page=3,trim=1.9cm 1.8cm 1.8cm 1.8cm, clip, width=1\textwidth]{SI.pdf}
\clearpage
\includegraphics[page=4,trim=1.9cm 1.8cm 1.8cm 1.8cm, clip, width=1\textwidth]{SI.pdf}
\clearpage
\includegraphics[page=5,trim=1.9cm 1.8cm 1.8cm 1.8cm, clip, width=1\textwidth]{SI.pdf}
\end{center}

\end{document}